\documentclass[12pt]{article}

\usepackage{paper2e}
\usepackage{mydefs2e}

\makeatletter
	\@addtoreset{equation}{section}%
\makeatother

\newcommand{\Li}{\hbox{\rm Li}}

\begin{document}

\begin{titlepage}
\preprint{KEK-TH-840\\ UMD-PP-03-011} 

\title{Almost No-Scale Supergravity}

\author{Markus A. Luty\footnote{E-mail: {\tt mluty@physics.umd.edu}}
\ \ Nobuchika Okada\footnote{Address starting September 2002:~Theory Group,
KEK, Tsukuba, Ibaraki 305-0801, Japan.\\
E-mail: {\tt okadan@post.kek.jp}}}

\address{Department of Physics, University of Maryland\\
College Park, Maryland 20742, USA}

\begin{abstract}
We construct an explicit 5-dimensional supergravity model that realizes the
`no scale' mechanism for supersymmetry breaking with no unstable moduli.
Supersymmetry is broken by a constant superpotential localized on a
brane, and the radion is stabilized by Casimir energy from supergravity
and massive hypermultiplets.
If the standard model gauge and matter fields are localized on a brane,
then visible sector supersymmetry breaking is dominated by gravity loops
and flavor-violating hypermultiplet loops,
and gaugino masses are smaller than scalar masses.
We present a realistic model in which the
the standard model gauge fields are partly localized.
In this model visible sector
supersymmetry breaking is naturally gaugino mediated, while masses of
the gravitino and gravitational moduli are larger than the weak scale.
\end{abstract}

\end{titlepage}

\noindent
Supersymmetry (SUSY) is arguably the most compelling solution to the
hierarchy problem, and SUSY breaking in extra dimensions arguably gives the
most attractive solution to the naturalness problems of SUSY
\cite{RS0,gMSB}.
In this Letter, we show that a very simple 5D model
can realize SUSY breaking of `no scale' type \cite{noscale},
with all moduli fields stabilized.
In this model the mass of the gravitino and bulk gravitational moduli are
much larger than the scale of supersymmetry breaking in the visible sector.
(5D warped compactifications with this feature were recently discussed in
\Ref{warpmod}.)
This elegantly solves the SUSY cosmological problems associated with
the gravitino and gravitational moduli.
The spectrum is that of gaugino-mediated \cite{gMSB} (or radion-mediated
\cite{RMSB}) SUSY breaking.

The present model consists of minimal 5-dimensional supergravity (SUGRA)
compactified on a $S^1 / Z_2$ orbifold with radius $r$.
We assume that the brane tensions on the orbifold fixed points
are small in units of the 5D Planck scale $M_5$, so that
the bulk metric is approximately flat.
We assume that there are constant superpotentials localized on one
or both orbifold fixed points.
This gives rise to a KK spectrum of spin-$\sfrac 32$ fermions \cite{BFZ}
\footnote{There is also a single massless spin-$\sfrac 12$ fermion,
the superpartner of the radion.}
\beq[gravKKmass]
m_{3/2}^{(n)} = \frac{\pm n + (\de_0 + \de_1) / 2\pi}{r},
\qquad
n = 0, 1, \ldots
\eeq
Here $\pm 1$ is the intrinsic orbifold parity of the mode, and
\beq[deltadef]
\de_{0,1} = 2 \tan^{-1} \frac{c_{0,1}}{2 M_5^3},
\eeq
where $c_{0,1}$ are the constant superpotentials localized on the two
branes and $M_5$ is the 5D Planck scale (normalized as defined below).
We will assume for simplicity that $c_1 = 0$, and $c = c_0 \ne 0$.

The graviton KK spectrum is not affected by the constant superpotentials,
so for $c / M_5^3 \ll 1$, the spectrum is nearly supersymmetric.
(We expect $c / M_5^3 \ll 1$
if $c$ arises from gaugino condensation on the brane.)
In this case the 4D effective field theory can be written as a supersymmetric
effective field theory with SUSY broken spontaneously by the $F$ term of the
radion.
The effective lagrangian is
\beq\bal
{\cal L}_4 = \myint d^4\th\, & \phi^\dagger \phi \left[
-3 M_5^3 (T + T^\dagger) + \De K(T^\dagger, T) \right]
\\
&
+ \left( \myint d^2\th\, \phi^3 \left[ c + \De W(T) \right]
+ \hc \right),
\eal\eeq
where $T = \pi r + \cdots$ is the radion chiral multiplet,
$\phi = 1 + \th^2 F_\phi$ is the conformal compensator of 4D SUGRA.
This effective theory follows directly from the formulation of 5D
SUGRA in terms of $\scr{N} = 1$ superfields given in \Ref{SUGRA5D}.
We have added a radion-dependent \Kahler potential and superpotential,
which requires additional 5D fields and interactions
to be discussed below.

If we neglect $\De W$ and $\De K$, we find that supersymmetry is broken
by
\beq[FT]
\avg{F_T} = \frac{c^*}{M_5^3},
\eeq
and the gravitino mass is
\beq
m_{3/2} = \frac{c}{M_4^2},
\eeq
where $M_4^2 = M_5^3 2 \pi r$ is the 4D Planck scale.
The potential vanishes identically, and $\avg{F_\phi} = 0$.
This is the `no scale' limit.
This limit will certainly not survive quantum corrections, and the radius
must be stabilized to obtain acceptable phenomenology.
Comparing \Eq{FT} with \Eqs{gravKKmass} and \eq{deltadef}, we see that
the 4D effective theory is valid if $\avg{F_T} \ll 1$.

We now include $\De W$ and $\De K$ as perturbations.
To linear order,
\beq[radpot1]
V = - \frac{|c|^2}{M_5^3} \De K_{T^\dagger T}
- \frac{1}{M_5^3} \left( c^* \De W_T + \hc \right),
\eeq
where $\De W_T = \partial (\De W) / \partial T$, {\it etc\/}.
Higher order terms are suppressed by additional powers of $M_5^3$.
There is no reason for the \Kahler and superpotential contributions to the
potential to be of the same size, so we expect one or the other to
dominate.
If the superpotential term dominates, it is easy to see that there is
no stable minimum at linear order in $\De W$.%
\footnote{The stabilization mechanism of \Ref{LS1} makes use of a
radius-dependent dynamical superpotential from bulk
gaugino condensation.
Consistent with the present analysis, this does not
lead to SUSY breaking of the no-scale type.}
If the \Kahler contribution dominates, there is a stable minimum
provided that $\De K_{T^\dagger T}$ has a local maximum.
We therefore look for stabilization mechanisms that give a nontrivial
\Kahler potential for the radion.

Assuming that the potential \Eq{radpot1} has a stable minimum, we can cancel
the cosmological constant by adding an additional source of SUSY breaking
on one of the branes.
This adds a positive constant to the \rhs of \Eq{radpot1}.
We avoid contact terms between the visible sector and the
SUSY breaking sector by assuming that these sectors are localized on
different branes.
We then obtain
\beq
\avg{F_\phi} = \frac{c^* \De K_{T^\dagger T}}{3 M_5^6}.
\eeq
This will give an anomaly-mediated contribution to visible SUSY breaking
\cite{AMSB}.
As long as
\beq[noscalecond]
\avg{\De K_{T^\dagger T}} \ll \frac{M_5^3}{2 \pi r},
\eeq
we have $\avg{F_\phi} \ll m_{3/2}$, and SUSY breaking masses
in the visible sector can be small compared to $m_{3/2}$.
We call such a model a `almost no-scale' model.
\Eq{noscalecond} is naturally satisfied if the characteristic mass scale of the 
stabilization dynamics is below $M_5$.

A natural candidate for \Kahler stabilization is Casimir energy \cite{Casimir}.
The gravitational contribution to the Casimir energy in this model
is \cite{BFZ}
\beq[gravCasimir]
V_{\rm grav} = (-4)
\frac{3 \zeta(3)}{8 \pi^2 L^4} |F_T|^2
+ \scr{O}(F_T^4),
\eeq
where $L = 2\pi r$.
We see that the gravitational contribution to the Casimir energy is
attractive, \ie 
favors small values of the radius.

The KK spectrum of a massive hypermultiplet 
in this theory can be worked out using the results of \Refs{SShyper}.
At each KK level there are 2 states with orbifold parity $+1$
and 2 states with orbifold parity $-1$:
\beq
\left( m_0^{(n)} \right)^2 =
\left( \frac{\pm n + F_T/ 2\pi }{r} \right)^2
+ m^2,
\quad
\left(  m_{1/2}^{(n)} \right)^2
= \left( \frac{n}{r} \right)^2 + m^2,
\quad
n = 0, 1, \ldots
\eeq
where $\pm 1$ is the intrinsic orbifold parity.
The Casimir energy is then (see \eg \Ref{PP})
\beq\bal
V_{\rm hyper} = (+2) \frac{3}{8\pi^2 L^4} \Bigl[ &
\sfrac 13 (m L)^2 \Li_1(e^{-m L})
\\
&
+ (m L) \Li_2(e^{-m L})
+ \Li_3(e^{-m L}) \Bigr] |F_T|^2
+ \scr{O}(F_T^4).
\eal\eeq
The asymptotic behavior is
\beq
V_{\rm hyper} \to
\begin{cases}
\displaystyle
(+2) \frac{3 \zeta(3)}{8 \pi^2 L^4} |F_T|^2
+ \scr{O}(F_T^4)
& for $L \ll 1/m$,
\\
\displaystyle
(+2) \frac{m^2 e^{-m L}}{8 \pi^2 L^2} |F_T|^2
+ \scr{O}(F_T^4)
& for $L \gg 1/m$.
\end{cases}
\eeq
The asymptotic behavior is easily understood physically.
For $L \ll 1/m$ the mass is negligible, and the Casimir energy goes like
$1/L^4$ on dimensional grounds.
The exponential suppression for $L \gg 1/m$ is the Yukawa
suppression of a massive scalar propagating over
distances of order $L$.

If there are 3 or more hypermultiplets, the
potential for $L \ll 1/m$ is dominated by the repulsive hypermultiplet
contribution, while for $L \gg 1/m$ it is dominated by the attractive
gravitational contribution.
There is therefore a minimum at $L \sim 1/m$.
The fact that massive modes can lead to Casimir stabilization was pointed
out by \Ref{PP}. 

The radion mass is $m_{\rm radion}^2 \sim \avg{F_T}^2 m^4 / (16\pi^2 M_4^2)$
and $K_{T^\dagger T} \sim \avg{F_T}^2 m^4 / (16\pi^2)$.
At the minimum the Casimir energy is negative and of order
$F_T^2 m^4 / (16\pi^2)$.
This contribution to the cosmological constant can be cancelled by an
additional source of supersymmetry breaking localized on one of the branes.
As long as $m \ll M_5$ this can be treated as
a perturbation on the analysis above.

We now discuss supersymmetry breaking in the visible sector.
As already mentioned, we localize the visible sector and the SUSY
breaking sector on different branes to avoid flavor-violating contact
terms between these sectors.
We must also consider possible flavor-violating contact terms arising
from the bulk hypermultiplets.
Tree-level contributions from the bulk hypermultiplet can be forbidden
by imposing a $Z_2$ symmetry on the hypermultiplet.
The leading couplings of the hypermultiplet to the visible and hidden
sectors that cannot be forbidden by symmetries are
\beq
\De\scr{L}_5 \sim \myint d^4\th \left[
\de(x^5) \frac{1}{M_5^3} Q^\dagger Q H^\dagger H
+ \de(x^5 - \pi r) \frac{1}{M_5^3} X^\dagger X H^\dagger H \right].
\eeq
These give rise to flavor-violating
contact terms in the 4D theory of order
\beq[badcontact]
\De\scr{L}_4 \sim \myint d^4\th\, \frac{1}{16\pi^2}\,
\frac{1}{M_4^4 L^2} Q^\dagger Q X^\dagger X.
\eeq
It is interesting to consider the possibility that SUSY breaking is
communicated to the visible sector by anomaly mediation.
However, for the Casimir stabilization considered above,
the SUSY breaking from \Eq{badcontact} is always larger than
the anomaly-mediated contribution.
To see this, we write
$\avg{\De K_{T^\dagger T}} \sim \ep m^4$
(where $\ep \sim 1/(16\pi^2)$ for Casimir stabilization)
and treat $L$ as independent of $m$.
Using $\avg{F_\phi} \sim \avg{F_T} \ep m^4 / M_5^3$
and $\avg{F_X}^2 \sim \avg{F_T}^2 \ep m^4$ we obtain
\beq
\frac{\De m^2_{\rm AMSB}}{\De m^2_{\rm loop}}
\sim \frac{\ep}{16\pi^2} (m L)^4.
\eeq
For Casimir stabilization the \rhs is of order $10^4$,
but anomaly mediation may dominate for other radius stabilization mechanisms
that give a flatter radion potential.
(Because the hypermultiplet loop contribution violates flavor, we need
$\De m^2_{\rm AMSB} \gsim 10^2 \De m^2_{\rm loop}$
to suppress flavor-changing neutral currents.)

We instead consider a different mechanism for transmitting SUSY breaking
to the visible sector.
We assume that the standard-model gauge multiplet is not completely localized,
but rather `leaks out' somewhat into the bulk.
$\avg{F_T} \ne 0$ then gives rise to a gaugino mass \cite{RMSB} with
a suppression factor due to the localization.

We can give an explicit realization of this scenario by assuming that 
brane-localized kinetic terms dominate over the bulk kinetic term for
the gauge zero modes \cite{DGS}.
We have
\beq
\frac{1}{g_4^2} = \frac{1}{g_{\rm bdy}^2} + \frac{L}{g_5^2}
\eeq
where $g_4$ is the 4D effective gauge coupling, $g_5$ is the 5D gauge coupling,
and $1/g_{\rm bdy}^2$ is the
coefficient of the brane-localized kinetic term.
In order for the brane-localized contribution to dominate, we must have
\beq
\frac{L}{g_5^2} \ll \frac{1}{g_{\rm bdy}^2} \sim \frac{1}{g_4^2} \sim 1.
\eeq
This requires small radius, while we require large radius for sequestering.
To see that there is a solution, we must be more careful about numerical
factors.

We estimate the size of unknown counterterms by assuming that bulk interactions
are strongly coupled at a scale cutoff $\La$ \cite{braneNDA}.
This gives
\beq
\La^3 \sim \ell_5 M_5^3,
\qquad
g_5^2 \sim \frac{\ell_5}{\La},
\eeq
where $\ell_5 = 24\pi^3$ is the 5D loop factor.
Assuming that massive flavor-violating bulk modes have masses of order
$\La$, sequestering requires $e^{-\La L} \lsim 10^{-3}$, so
$\La L \gsim 7$ is sufficient.
We then have
\beq
\frac{L}{g_5^2} \gsim \frac{\La L}{\ell_5} \sim 10^{-2}.
\eeq

The gaugino mass is given by
\beq
m_{1/2} \sim \frac{\avg{F_T}}{g_5^2},
\eeq
so that
\beq
\frac{m_{3/2}}{m_{1/2}} \sim \frac{\ell_5}{\La L} \lsim 100.
\eeq
We see that the gravitino mass can be $\sim 100$ times larger than
the weak scale in this model.
This is plausibly enough so that $m_{3/2} > 60\TeV$, in which case
the gravitino decays early enough to avoid problems with nucleosynthesis
\cite{GR}.

To avoid flavor-changing neutral currents, we must have
$m_{1/2}^2 \gsim 10^3 \De m_{\rm scalar}^2$,
where $\De m_{\rm scalar}^2$ is the flavor-violating
scalar mass contribution from operators
of the form \Eq{badcontact}.%
\footnote{
Gravity loops give flavor-blind contributions to the soft masses 
that are also of order $\De m_{\rm scalar}^2$.}
This is satisfied as long as $\La L \gsim m^2/ M_4^2$, which is always
satisfied.
We conclude that this gives 
an interesting model in which the spectrum is gaugino-mediated,
and yet the gravitino is heavy.

We briefly comment on the radiative stability of this model.
4D gravity loops are cut off in the ultraviolet for momenta $p_4 \sim 1/r$,
where the extra dimension becomes important.
In the 5D theory, gravity loops that contribute to SUSY breaking must
connect the two branes, and the contribution from $p_4 \gg 1/L$ is 
therefore suppressed by $e^{-p_4 L}$.
The extra dimension acts as a `low' cutoff (below $M_4$) that makes this
scenario radiatively stable.

If the gaugino is more strongly localized, we can get a larger value for
the ratio $m_{3/2} / m_{1/2}$.
The consistency of such a scenario with local 5D supersymmetry is
strongly suggested by the fact that we can write an
effective theory where the gauge fields are completely
localized on an effective-theory brane.
An additional hint comes from soliton solutions in higher-dimensional
supergravity, which give rise to solutions
where $U(1)$ gauge fields are localized on a brane \cite{KM}.

In a more non-minimal model, there may be additional gravitational
moduli, light scalar fields that interact only by Planck-suppressed
operators, and which have an exactly flat potential in the SUSY limit.
Such fields will also get a mass of order $m_{3/2}$ from contact
terms with the hidden sector.
If $m_{\rm moduli} \gsim 100\TeV$, this solves the Polonyi problem
associated with these moduli.

In conclusion, we have presented a simple 5D model that realizes the no-scale
mechanism for supersymmetry breaking, with the radion stabilized by Casimir
forces.
The superpartner spectrum in this model is gaugino mediated, and the
main new feature is that the mass of the gravitino and gravitational
moduli are heavier than the weak scale,
eliminating cosmological problems.
More generally, the lesson from this paper (and from \Ref{warpmod}) is
that the gravitino mass is controlled by UV physics that is independent
from that which gives rise to visible sector SUSY breaking.

\section*{Acknowledgements}
We thank M. Schmaltz for discussions.
This work was supported by NSF grant PHY-0099544.

\newpage

\end{document}